\documentclass[prd,superscriptaddress,preprintnumbers,nofootinbib,showkeys,notitlepage,twocolumn
]{revtex4-1}
\usepackage{graphicx,amssymb,amsmath,color}
\usepackage{hyperref}
\usepackage{multirow}
\usepackage{ulem}

\newcommand{\beq}{\begin{equation}}
\newcommand{\eeq}{\end{equation}}
\def\bea{\begin{eqnarray}}
\def\eea{\end{eqnarray}}
\newcommand{\bei}{\begin{itemize}}
\newcommand{\eei}{\end{itemize}}

\newcommand{\Fig}[1]{Fig.~\ref{#1}}
\newcommand{\Eq}[1]{Eq.~(\ref{#1})}
\newcommand{\Sec}[1]{Sec.~\ref{#1}}
\newcommand{\App}[1]{Appendix~\ref{#1}}

\def\={\,=\,}
\def\+{\,+\,}
\def\-{\,-\,}
\def\Msun{M_{\odot}}

\begin{document}

\title{
Constraining the gravitational coupling of axion dark matter at LIGO
}

\author{Sunghoon Jung}
\email{sunghoonj@snu.ac.kr}
\affiliation{Center for Theoretical Physics, Department of Physics and Astronomy, Seoul National University, Seoul 08826, Korea}

\author{TaeHun Kim}
\email{gimthcha@snu.ac.kr}
\affiliation{Center for Theoretical Physics, Department of Physics and Astronomy, Seoul National University, Seoul 08826, Korea}

\author{Jiro Soda}
\email{jiro@phys.sci.kobe-u.ac.jp}
\affiliation{Department of Physics, Kobe University, Kobe 657-8501, Japan}

\author{Yuko Urakawa}
\email{urakawa.yuko@h.mbox.nagoya-u.ac.jp}
\affiliation{Fakult{\"u}t f{\"u}r Physik, Universit{\"a}t Bielefeld, Bielefeld 33501, Germany}
\affiliation{Department of Physics and Astrophysics, Nagoya University, Chikusa, Nagoya 464-8602, Japan}


\begin{abstract}

The axion-gravity Chern-Simons coupling is well motivated but is relatively weakly constrained, partly due to difficult measurements of gravity. We study the sensitivity of LIGO measurements of chirping gravitational waves (GWs) on such coupling. When the frequency of the propagating GW matches with that of the coherent oscillation of axion dark matter field, the decay of axions into gravitons can be stimulated, resonantly enhancing the GW. Such a resonance peak can be detected at LIGO as a deviation from the chirping waveform. Since all observed GWs will undergo similar resonant enhancement from the Milky-Way (MW) axion halo, LIGO O1+O2 observations can potentially provide the strongest constraint on the coupling, at least for the axion mass $m_a = 5 \times 10^{-13} - 5 \times 10^{-12}$ eV. Along the course, we also emphasize the relevance of the finite coherence of axion fields and the ansatz separating forward and backward propagations of GWs. As a result, the parity violation of the Chern-Simons coupling is not observable from chirping GWs.

\end{abstract}

\preprint{KOBE-COSMO-20-03}

\maketitle

\tableofcontents

\section{Introduction}

The axion is an important candidate of dark matter. 
Axions are not restricted to the QCD axion, but a variety of axions are predicted from stringy setups~\cite{Arvanitaki:2009fg}. They are very light pseudo-scalar particles coupling to Chern-Simons terms of some gauge fields $F\widetilde{F}$. Combined with proper cosmological histories, a wide range of axions can be a full dark matter candidate (see e.g. \cite{Graham:2018jyp}).

However, the axion is very elusive as it couples to standard model particles very weakly, suppressed by its large decay constant $f_a$. Thus, usual direct detection experiments are not sensitive to the axion. A whole new varieties of axion detection experiments and astrophysical probes have been proposed, mainly based on its lightness (due to the pseudo Goldstone nature) and the coherent oscillation (due to the non-relativistic dark matter nature)~\cite{Graham:2011qk}.
They can constrain the axion couplings to photons and electrons, for example through supernova cooling, oscillating electric dipole moments, birefringence of pulsars, quasars, and cosmic microwave background (CMB), and the mixing with the photon inside electron plasma. We refer to \cite{Marsh:2015xka} for reviews. 

But the axion-gravity coupling is relatively weakly constrained. Similarly to the axion-photon $aF \widetilde{F}$ coupling, the axion-gravity Chern-Simons coupling can be generically produced \cite{Choi:1999zy, Kim:2016ncr}. Whenever there is a gravitational anomaly, there must exist an associated axion coupling to the gravity. The latest bound on the axion-gravity coupling $\ell \lesssim 10^8$ km (\Eq{eq:elldef}) comes from the measurement of frame-dragging effects around the Earth by Gravity Probe B~\cite{AliHaimoud:2011fw}.

In the meantime, the chirping gravitational wave (GW) from a binary merger arises as a new tool to probe the Universe. Since it has a well predicted waveform chirping in time and frequency domains in a particular way, even small perturbations to the chirping can be confidently detected. Example studies with dark matter perturbations are \cite{Giudice:2016zpa,Jung:2017flg,Lai:2018rto,Christian:2018vsi,Dai:2018enj,Jung:2018kde}, one of which is probing coherently oscillating light dark matter around binary mergers~\cite{Choi:2018axi}.

In this paper, we study how the chirping GW can be perturbed by the coherent axion field as the GW propagates through it. Although the gravitational perturbation is usually very small, a resonant phenomenon can occur when the GW frequency matches with the axion Compton frequency. The resulting signal is a resonance peak in the frequency spectrum.

The resonant phenomenon on the electromagnetic (EM) wave has been studied with various observables. For example, the modification of the EM wave propagating through the coherent axion dark matter field can produce a sharp resonance peak in the frequency spectrum \cite{Yoshida:2017ehj, Arza:2018dcy, Caputo:2018ljp, Caputo:2018vmy, Rosa:2017ury} or can even produce an echo coming back to us \cite{Arza:2019nta}. The resonance can also destabilize axion structures~\cite{Hertzberg:2018zte, Wang:2020zur}, possibly leaving some signals in the background or producing an explosive burst \cite{Tkachev:2014dpa}. 

On the other hand, the GW resonance from the coherent dark matter field has not been studied in detail, even though the axion-gravity coupling is well motivated too. Up to our knowledge, the GW resonance was first studied in \cite{Yoshida:2017cjl}, but it lacks detailed analysis of realistic observables. Our work aims at providing an elaborate analysis for the GW resonance and using it to probe axion-gravity couplings with the LIGO. We will mainly focus on the modification of chirping GWs, but will discuss the instability of axion substructures too. Readers may also refer to \cite{Chu:2020iil} for other non-resonant GW observables and \cite{Kitajima:2018zco} for axion-generated GW background.

Our work also improves upon the previous works on the resonance in that we correctly include the finite spatial coherence of the axion field and separate the forward and backward waves. Although similar analyses have been done for EM waves in \cite{Arza:2018dcy, Arza:2019nta}, the former ignored the spatial coherence while the latter did not discuss the forward wave. Both treatments are crucial in the LIGO observation, and the absence of parity violation observables is one remarkable consequence.

The paper is structured as follow. We start with a summary of main points and physics of the paper in \Sec{sec:summary}. We derive and solve wave equations in \Sec{sec:sol}, introduce our axion signals on the chirping GW in \Sec{sec:signal}, and present LIGO bounds and prospects in \Sec{sec:bound}. We provide further details on the resonance with various viewpoints in \Sec{sec:resonance}, and discuss interesting findings on the absence of parity violation in \Sec{sec:parityv}. Then we conclude in \Sec{sec:conclusion}.

\section{Overview} \label{sec:summary}

We consider the MW axion halo, which is a highly coherent superposition of axion waves, with a long spatial coherence $\sim 1/ m_a \Delta v$ (with a small velocity dispersion $\Delta v$) and a much longer temporal coherence (longer than the duration of the chirping GW in the LIGO band). 
The long coherence stems from the non-relativistic nature ($v\sim \Delta v \ll 1$) of the axion dark matter.

The coherent (temporal) oscillation can induce resonant enhancement of the chirping GW, when the GW frequency matches with the axion Compton frequency. 
Since the waveform of the chirping GW is very well predicted, the resonance peak can be detected. It can be further distinguished from accidental noise because all observed GWs will experience a similar phenomenon from the MW axion halo. We found that the correlation of all 11 GW observations at LIGO O1+O2 can provide one of the strongest constraints on the axion-gravity coupling.

The resonant phenomenon is essentially the stimulated decay of axions, although we treat those waves classically. We present several analyses to make sense of the particle-like interpretation of the solution of wave equations that we actually obtain and use.

\medskip
Other remarkable technical points:

The finite (spatial) coherence does impact the signal. Not only does it reduce the enhancement, but it also broadens the frequency width and induces finite time-duration of the resonance peak.

We distinguish forward and backward-going GWs generated from the propagation through an axion halo. First, only forward waves from a binary merger will be observed. Second, backward waves must be generated by the energy-momentum conservation, if forward waves are to be enhanced. Last, mostly only forward and backward waves are generated, which can be understood from a symmetry consideration well inside a halo.

The distinction of forward and backward waves leads to different observable relations of the parity violation. In our case, the parity violation exists only on backward waves, hence not observable. But in existing studies, parity violation was observable because non-resonant regime was considered and/or stochastic waves were considered where the forward/backward distinction is not possible.

\section{Propagation through coherent axions}  \label{sec:sol}

We solve coupled wave equations between axion fields and GWs by using an ansatz suitable for the propagating GW.  
Then we discuss the solution near a resonance regime with small enhancement.

\subsection{Coupled wave equations}

The gravitational Chern-Simons  Lagrangian $\mathcal{L} = \frac{\alpha}{4} a R \tilde{R}$ gives the linearized action in the flat background as (ignoring the cosmic expansion)
\begin{eqnarray}
&&S_{\rm EH} + S_{\rm CS} \nonumber \\
&&= \frac{\kappa}{4} \int d^4 x \left[h^i{}_{j,t}h^j{}_{i,t}-h^i{}_{j,k}h^j{}_{i,}{}^k \right. \nonumber \\
&& \qquad \qquad \quad \ \left. - \frac{\alpha}{\kappa} \dot{a} \epsilon^{ijk}\left(h^l{}_{i,t}h_{kl,jt}-h^l{}_{i,}{}^m h_{kl,mj}\right) \right],
\end{eqnarray}
where $\alpha$ is the gravitational Chern-Simons  coupling constant, $h_{ij}$ is the metric perturbation and $\kappa = 1/16\pi G$. Varying this action with respect to the metric perturbation $h_{ij}$ gives the wave equation \cite{Alexander:2004wk, Alexander:2009tp}
\begin{equation}
\left(\partial_t^2-\vec{\nabla}^2\right)h^j{}_i  \=  \frac{\alpha}{\kappa} \epsilon^{lkj}\left(\ddot{a}h_{ki,lt}+\dot{a}h_{ki,ltt}-\dot{a}h_{ki,m}{}^m{}_l\right). \label{eq:waveeq}
\end{equation}

We approximate the axion field $a$ to have only time dependence through its Compton oscillation (spatially homogeneous)
\begin{equation}
a(t)  \=  \frac{a_0}{2} e^{-i m_a t} + \text{c.c.}, \label{eq:a}
\end{equation}
where $a_0$ is the complex amplitude (containing the initial phase information).
Axions are non-relativistic (a dark matter candidate) so their small kinetic energy contribution to the Compton frequency is neglected. The amplitude $a_0$ (hence, the energy density) is assumed to be constant in time, as the energy density of the axion field is much larger than that of the chirping GW (\Sec{sec:energyconservation}). For a more realistic axion halo spatial profile, see \Sec{sec:halo}.

\medskip

To solve \Eq{eq:waveeq} for the propagating GW in a finite axion halo, we introduce an ansatz for $h_{ij}$ considering\footnote{In \App{app:Mathieu}, we present another approach of solving the wave equation, giving the same result.}
\begin{enumerate}
\item Plane waves propagating in the $\hat{z}$ direction.
\item Backward wave. The conservation of momentum enforces the generation of backward propagating waves when the forward wave is enhanced\footnote{From a symmetry consideration, the generation of only forward and backward waves must be true, at least well inside a finite halo. But there can be slight leakage over all directions near the boundary of a halo or a coherent patch, although the boundary still varies smoothly over a large scale. We ignore the leakage.}. We will distinguish forward and backward waves, in order to describe forward propagating GWs that we eventually observe. This leads to different observable relations from previous works; see \Sec{sec:parityv}. 
\item Circular polarization. The Levi-Civita tensor $\epsilon^{ijk}$ mixes the $+$ and $\times$ polarizations, while right handed (R) and left handed (L) circular helicities are decoupled. 
\end{enumerate}
These conditions give the following ansatz (similarly to the photon ansatz introduced in \cite{Arza:2018dcy} but in the circular polarization basis):
\begin{equation}
h_{ij}(z, t)  \=  h_{ij}^{(R)}(z,t) \+ h_{ij}^{(L)}(z,t),
\end{equation}
where each helicity mode is expressed as\footnote{The ansatz with $e^{- i \frac{m_a}{2} t}$ rather than $e^{-i \omega t}$ is more convenient to solve the equation, but we will check this ansatz gives the correct dispersion relation; see below \Eq{eq:psi}.}
\begin{eqnarray}
h_{ij}^{(s)}(z, t) &=& \hat{e}_{ij}^{(s)} h_F^{(s)}(t) e^{i\left(kz-\frac{m_a}{2}t\right)} \nonumber \\ &&-  i\hat{e}_{ij}^{(\bar{s})} h_B^{(s)}(t) e^{i\left(-kz-\frac{m_a}{2}t\right)} \+ \text{c.c.},  \label{eq:ansatz}
\end{eqnarray}
where $h_F$ and $h_B$ are complex amplitudes for the forward and the backward waves, with the superscript $s = L,R$ denotes helicity and $\bar{s}$ refers to the opposite to $s$. The polarization tensor $\hat{e}_{ij}^{(s)}$ is defined with respect to the direction of $+\hat{z}$ propagation. 
 
Applying Eqs. (\ref{eq:a}) -- (\ref{eq:ansatz}) into the wave equation (\ref{eq:waveeq}) (with $|\ddot{h}/ \dot{h}| \ll m_a$) gives coupled first-order differential equations for forward and backward waves. They become decoupled in the second order equations as 
\begin{eqnarray}
\ddot{h}_{F/B}^{(s)}(t) &=& \left(\frac{m_a}{2}\right)^2 \left(2\pi G \ell^4 m_a^4 \frac{|a_0|^2}{4} - \epsilon^2 \right) h_{F/B}^{(s)}(t) \nonumber \\ 
&=& \left(\frac{m_a}{2}\right)^2 \left(\gamma^2 - \epsilon^2 \right) h_{F/B}^{(s)}(t) \nonumber \\
&=& \mu^2(\epsilon) h_{F/B}^{(s)}(t), \label{eq:2ndeq}
\end{eqnarray}
where 
\begin{equation}
\ell^2 \, \equiv\, \alpha / \sqrt{\kappa / 2}   \label{eq:elldef}
\end{equation}
is the coupling parameter $\ell$ that we use to describe the axion-gravity coupling \cite{Yoshida:2017cjl, Okounkova:2017yby},
\begin{equation}
\epsilon \, \equiv \, \frac{k - m_a/2}{m_a/2}
\end{equation}
is the fractional deviation of $k$ from the resonance frequency $m_a/2$, and a useful dimensionless combination of parameters is
\bea
\gamma & \equiv & \sqrt{2\pi G} l^2 m_a^2 \frac{|a_0|}{2} \label{eq:gamma} \\
&=& 5.7 \times 10^{-9}  \label{eq:gamma0} \\
&&\times \left( \frac{\ell}{10^8 \, {\rm km}} \right)^2 \left( \frac{m_a}{10^{-13} \,{\rm eV}} \right) \left(\frac{\rho_a}{0.3 \,{\rm GeV/cm}^3} \right)^{1/2}, \nonumber
\eea
where $m_a^2 |a_0|^2 /2 = \rho_a$ is used. For $m_a = 10^{-13} \sim 10^{-10}$ eV relevant to the LIGO band, $\gamma \ll 1$ for the most range of currently allowed coupling and density. Thus, the enhancement rate \begin{equation}
\mu \, \equiv \, \frac{m_a}{2} \sqrt{\gamma^2 - \epsilon^2} \label{eq:mu}
\end{equation}
will be assumed to be small throughout the paper. The origin of the name is clear from \Eq{eq:2ndeq} which describes exponential enhancement when $\mu$ is real. These parameters will be used widely in our phenomenology study.

\subsection{Solution for finite propagation}  \label{sec:solution}

The solutions of the wave equation \Eq{eq:2ndeq} can be expressed in terms of initial values $h_F^{(s)}(0)$ and $h_B^{(s)}(0)$ as
\begin{eqnarray}
h_{F/B}^{(s)}(t) &=& h_{F/B}^{(s)}(0) \cosh(\mu t) \nonumber \\ 
&& + \left[i \lambda^{(s)} \frac{\gamma e^{i\phi_0}}{\sqrt{\gamma^2-\epsilon^2}} \left(h_{B/F}^{(s)}(0) \right)^{*} \right. \nonumber \\ 
&& \qquad \ \left. -i \frac{\epsilon}{\sqrt{\gamma^2-\epsilon^2}} h_{F/B}^{(s)}(0) \right] \sinh(\mu t), \qquad  \label{eq:gensol}
\end{eqnarray}
where $\lambda^{(R/L)} = +1/-1$ and $\phi_0$ denotes the phase part of the axion amplitude as $a_0 = |a_0| e^{i\phi_0}$. 

The initial condition relevant to the forward-propagating chirping GW is $h_B^{(s)}(0)=0$. Then, the solutions are 
\begin{subequations}
\begin{eqnarray}
h_F^{(s)}(t) &=& h_F^{(s)}(0) \left[\cosh(\mu t) - i \frac{\epsilon}{\sqrt{\gamma^2-\epsilon^2}} \sinh(\mu t) \right] \qquad \label{eq:sol} \\
h_B^{(s)}(t) &=& i \lambda^{(s)} \frac{\gamma e^{i\phi_0}}{\sqrt{\gamma^2-\epsilon^2}} \left(h_F^{(s)}(0)\right)^{*} \sinh(\mu t). \qquad \label{eq:solback}
\end{eqnarray}
\end{subequations}
These GW solutions are of the same form as those of electromagnetic(EM) waves in \cite{Arza:2018dcy, Wang:2020zur}, even though the wave equations are different. These solutions are valid for both real and complex $\mu$'s. We hereafter focus only on the forward wave as it is what we observe from binary mergers. This is overlooked in the previous GW work \cite{Yoshida:2017cjl}; see \Sec{sec:parityv} for observational implications.

\medskip

Now consider finite propagation of GW with $\mu t \ll 1$ in \Eq{eq:sol}, where $t$ is the propagation time. This limit will be relevant to the finite coherent axion patch. We first express \Eq{eq:sol} in the polar form
\begin{equation}
h_F^{(s)}(t)  = h_F^{(s)}(0) \times F(t) \times e^{-i \psi(t)},  \label{eq:defF}
\end{equation}
and express $F(t)$ and $\psi(t)$ up to their lowest order axion contributions under $\mu t \ll 1$ ($\gamma, \epsilon \ll 1$ is always assumed). They are  
\begin{eqnarray}
F(t) & \approx & 1+\frac{\gamma^2}{2} \left(\frac{m_a}{2} t\right)^2 \text{sinc}^2 \left(\frac{m_a}{2} \epsilon  t \right)\nonumber \\
&\equiv& 1+ \delta(\epsilon, t), \label{eq:F}
\end{eqnarray}
where $\delta \ll 1$ by $\mu t\ll1$ and
\begin{equation}
\psi(t) \approx \frac{m_a}{2}\epsilon t \left\{1+\frac{1}{2}\left[\text{sinc}\left(m_a\epsilon t\right) -1 \right] \left(\frac{\gamma}{\epsilon}\right)^2 \right\}. \label{eq:psi}
\end{equation}
The leading term in the phase $\psi$, combined with the phase of the ansatz in \Eq{eq:ansatz}, gives the phase velocity equal to the speed of light: $(m_a/2)t + (m_a/2)\epsilon t = k t$ so that $\omega = k$. Thus, the second term of \Eq{eq:psi} gives the correction to the dispersion relation as will be discussed in \Sec{sec:timedelay}. Hereafter, we no longer distinguish the wave number $k$ and the angular frequency $\omega$ in the leading order. Similarly, for $F(t)$ in \Eq{eq:F}, the 1 refers to the original wave and the second term $\delta(\epsilon,t)$ is the enhancement due to the stimulated axion decay. 

The resonance shape described by \Eq{eq:F} is different from the naive expectation from \Eq{eq:2ndeq}. This is due to the finite propagation time, or equivalently the finite coherence of the axion field. In the next \Sec{sec:signal1}, we discuss physical properties of these solutions with finite propagation time, in comparison to those with infinite propagation.

\section{Signal}  \label{sec:signal}

We introduce two kinds of axion signals, main one in \Sec{sec:signal1} and another in \Sec{sec:signal2}. In the last two subsections, we discuss how to calculate them from the propagation through multiple coherent axion patches of a galactic halo. 

\subsection{Signal 1: Resonance with finite coherence} \label{sec:signal1}

One may use the wave equation in \Eq{eq:2ndeq} to describe an exponential growth when $\mu$ is real for $\frac{m_a}{2} (1-\gamma) \,<\, k \,<\, \frac{m_a}{2}(1+\gamma)$ from \Eq{eq:mu}. The width $m_a \gamma$ is very narrow (see \Eq{eq:gamma0}) so that $k \approx m_a/2$. This relation is consistent with the particle interpretation of the phenomenon as a stimulated axion decay into two gravitons.  Thus, the growth is also called the `resonant enhancement'. The maximum enhancement rate from \Eq{eq:mu} $\mu_{\rm max} = m_a \gamma/2$ is also determined by $\gamma$.

\medskip

However, the finite coherence of the axion field makes important modifications on the resonance. First, the resonance width is broadened, not simply given by $m_a \gamma$ as above. The resonance in each coherent patch is given by \Eq{eq:F} with substituting $t$ by the coherent patch size $L_{\rm coh} \sim 1/m_a \Delta v \sim 1/m_a v$ ($\Delta v \sim v$ is the velocity dispersion of axions; see \Sec{sec:halo} for more details),
\begin{eqnarray}
F_{\rm patch} &=& 1+\delta_{\rm patch} (f) \nonumber \\
&=& 1+\frac{\gamma^2}{2} \left(\frac{1}{2 \Delta v} \right)^2 \text{sinc}^2 \left(\frac{\epsilon}{2 \Delta v} \right), \label{eq:Fpatch}
\end{eqnarray}
where $\delta_{\rm patch} \ll 1$ describes the enhancement at each patch. The frequency width of the enhancement is given by the central peak of the sinc function: $-2 \Delta v \lesssim \epsilon \lesssim 2 \Delta v$, corresponding to $m_a/2 - m_a \Delta v \lesssim \omega_{\rm GW} \lesssim m_a/2 + m_a \Delta v$ yielding the peak width $\sim 2 m_a \Delta v$. This is different from the estimation above, where the width was determined by $\gamma$. The modification can be understood from two perspectives. First, the length of the patch is $1/m_a \Delta v$, so each patch cannot have a frequency resolution better than $m_a \Delta v$. This determines the resonance width. Another point of view is that axions have the velocity dispersion $\Delta v$, so that the observed resonance width is Doppler broadened by fractionally $\Delta v$. These two perspectives are essentially the same, since the patch size is determined by the velocity dispersion. 

In addition, the broadened frequency width $m_a \Delta v$ implies the time duration $1/m_a \Delta v$ of the resonance, related by the Fourier transform. The time duration equals to the size of a coherent patch $L_{\rm coh}$, hence the time taken for a GW to pass one patch. Such a long duration, combined with the time-delay of a resonance, may affect detection methods as will be discussed in \Sec{sec:timedelay}.

\begin{figure}[t] 
\centering
\includegraphics[width=0.48\textwidth]{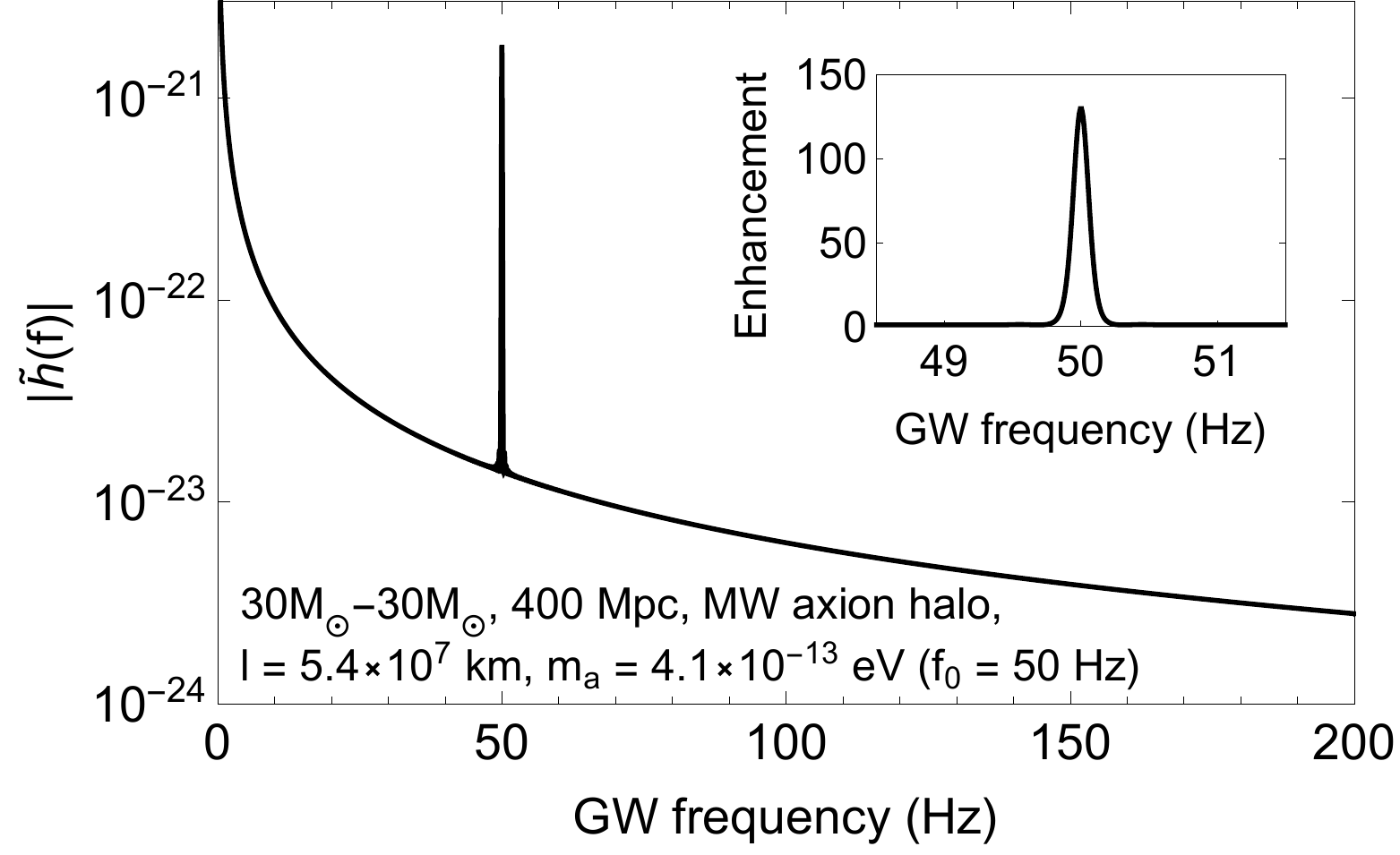}
\caption{An example axion signal in the chirping GW spectrum. The sharp peak is at the resonance frequency $f_0 = 50 \ \text{Hz}$ of the stimulated decay of axions with the mass $m_a = 4.1\times 10^{-13} \ \text{eV}$. The inset shows the total enhancement from the propagation through a 100 kpc axion halo. More details are as follow. The chirping GW is generated from 30-30 $\Msun$ binaries at $D=400$ Mpc. The effect of multiple coherent patches are maximally multiplied (\Sec{sec:addup}), and the resonance broadening is taken into account. The coupling strength, $\ell = 5.4 \times 10^7$ km, is chosen to produce a barely detectable peak (see \Sec{sec:criteria} and \Fig{fig:bounds}) showing that our detection criteria can be conservative. 
} 
\label{fig:enhancedGW}
\end{figure}

\medskip

Our main signal is a narrow resonance peak in the chirping GW frequency spectrum, produced by the resonant enhancement. We show an example signal in \Fig{fig:enhancedGW}, which results from the propagation through a 100 kpc axion halo consisting of many smaller coherent patches. Although the exact resonance shape depends on finite coherence and effects from multiple patches in \Sec{sec:addup}, the basic properties are as discussed above: the narrow peak at $k \simeq m_a /2$ and peak width and height determined largely by $\gamma$ and $\Delta v$.

Since the chirping waveform  from binary mergers is very well predicted and  does not usually accompany such a sharp peak, the absence of such a peak in the LIGO observations can constrain the resonance enhancement. The peak can be confused with instrumental noise which also often appears as a sharp peak. But since all GWs arriving at us will experience similar enhancements due to MW axion clouds, one can gain confidence by correlating all observed GWs in the frequency and time domains simultaneously. Therefore, if a peak is observed in one GW, a peak with similar properties (frequency, timing, and amplitude) must be observed in all GWs. We quantitatively study this signal with LIGO capability in \Sec{sec:results}.

\subsection{Signal 2: Explosion} \label{sec:signal2}

Another constraint comes from the existence of certain dark matter substructures. If $\mu$ is too large, the stimulation becomes quicker and quicker so that a coherent axion patch becomes unstable and decays almost entirely into GWs. Such happens when~\cite{Hertzberg:2018zte}
\beq
\mu_{\rm max} L_{\rm coh}  \= \frac{\gamma}{2 \Delta v} \,>\, 1 \qquad \textrm{(explosion),}  \label{eq:explosion}
\eeq
which essentially means that the enhancement rate $\mu$ is larger than the passing time within a coherent patch $L_{\rm coh} = 1/(m_a \Delta v)$ in \Eq{eq:cohpatch}. Thus, this may happen for small enough $\Delta v$ (long enough coherence) and high enough density $\rho_a$.

If there existed such substructures that could explode (satisfying the above condition), they must have almost disappeared by today because there are background photons and GWs everywhere with essentially any frequencies. Produced photons and GWs might have been dissipated enough or became a part of the stochastic backgrounds so that they might not be observable today. Instead, it is the observation of certain dark matter substructures surviving today which can impose an upper limit on the Chern-Simons coupling. 

The observed dark matter substructures with possibly the largest enhancement rate are likely dwarf galaxies. They have small velocity dispersion $\Delta v = {\cal O}(1-10)$ km $\simeq 10^{-5}$ (thus, the long coherence length) and large dark matter density $\rho \simeq 10^3 \times 0.3\, {\rm GeV/cm}^3$ at its central core~\cite{Tulin:2017ara} albeit some uncertainties.
We conservatively use these values to estimate the upper bound on the Chern-Simons coupling, from the existence of  dwarf galaxies; in any case, the bound on the coupling $\ell$ is not so sensitive to the density as it scales with $\rho_a^{1/4}$ in \Eq{eq:gamma0}. As shown by the red solid in \Fig{fig:bounds}, this constraint is weaker than that from the resonance peak. Also, the approximation with small $\mu t \ll 1$ for the MW axion halo is thus good.

\medskip
As an aside, which axion substructures could lead to explosion? The axion minicluster~\cite{Kolb:1993zz} has long coherence and high density. Virialized within its Jeans length, the axion minicluster has the Jeans length~\cite{Marsh:2015xka} 
\beq
r_J \= \frac{2\pi}{(16\pi G \rho_{\rm mc})^{1/4} m_a^{1/2}} \,\sim\, \frac{1}{m_a \Delta v},
\eeq
which is of order of the de Broglie wavelength. Here, $\rho_{\rm mc}$ is the density of a minicluster. Thus, the explosion $\mu_{\rm max} L_{\rm coh} \simeq \mu_{\rm max} r_J \gtrsim 1$ (\Eq{eq:explosion}) happens when
\bea
\ell \, &\gtrsim\,& (7.3\times 10^{6} \ \text{km}) \nonumber \\ &&\times \left(\frac{m_a}{10^{-12}\text{eV}}\right)^{-3/4}  \left(\frac{\rho_{\rm mc}}{0.3 \ \text{GeV/cm}^3}\right)^{-1/8}. \qquad
\eea
As expected, this value of $\ell$ is much smaller than the bounds coming from dwarf galaxies and resonances (cf. \Fig{fig:bounds}). Although this estimate can be subject to small gravitational redshifts due to the minicluster itself and axion self interactions~\cite{Wang:2020zur}, we conclude that axion miniclusters are irrelevant to our work. If miniclusters had existed, they would have almost disappeared by today by explosion, or the axion coupling is too weak to be probed by any methods.

\subsection{Modeling an axion halo with multiple coherent patches} \label{sec:halo}

A realistic axion halo is not infinitely coherent. The coherence property varies among axion dark substructures. As discussed in \Sec{sec:signal2}, it is good enough to consider an axion halo without miniclusters; such a scenario is motivated by the misalignment production mechanism~\cite{Preskill:1982cy,Abbott:1982af,Dine:1982ah}. 

Such axion halo is virialized with the Milky-Way (MW) whose total mass is $\sim 10^{12} \Msun$ in a radius of 100 kpc. The virial velocity $v \sim 10^{-3}$ with the Maxwellian dispersion $\Delta v \sim v \sim 10^{-3}$ leads to the superposition of axion fields
\beq
a (x,t) \= a_0 \cos (k x - \omega t + \phi)
\eeq
with the long spatial coherence length $1/m_a \Delta v \sim 1/m_a v \sim 10^3 / m_a$ (with $k=m_a v$)~\cite{Graham:2011qk}. This length is the size of a coherent patch 
\beq
L_{\rm coh} \= \frac{1}{m_a \Delta v}, \label{eq:cohpatch}
\eeq 
in that a halo has a spatially oscillating profile with the oscillation length scale of $L_{\rm coh}$. This is essentially the random-walk superposition of $N$ axion waves (with random phase) which leads to the total amplitude $|a_0| \propto \sqrt{N}$ consistent with the energy density given by $\rho_a = m_a^2 |a_0|^2 /2$ (see \Eq{eq:rhoave} for the value of $\rho_a$). In addition, $\omega = m_a (1+v^2/2)$ so that the temporal coherence is broken only after a long time $1/m_a \Delta v^2 \sim 10^6/m_a$, much longer than the GW propagation time in each coherent patch. Thus, we ignore the temporal incoherence while taking into account the spatial incoherence. 

Thus, an axion halo is composed of many smaller patches of sizes about the coherence length. As the GW propagates through an axion halo, it passes through multiple coherent patches. As the resonant effect grows only within a coherent patch, the total enhancement will be the sum of the individual patch's effect. We discuss how to sum them up in \Sec{sec:addup}.

In \Sec{sec:sol}, we have solved wave equations by assuming the spatially homogeneous and infinite axion field. We apply this solution to each coherent patch, which is actually of finite size and spatially varying.  In effect for simplicity, we approximate each coherent patch as a Heaviside profile with the length $L_{\rm coh}$  and the amplitude satisfying $\rho_a = m_a^2 |a_0|^2 /2$. The solution is thus good enough well inside the patch, but our calculation does not include the entrance and exit of GWs through the boundary of a patch. Nevertheless, this approximation can still capture the main physics of the phenomenon. We defer more accurate calculations to the future.

\subsection{Summing effects from multiple patches} \label{sec:addup}

A dark matter halo in a galaxy consists of many smaller coherent patches. The resonant enhancement occurs only within a coherent patch. Therefore, we need to sum the effects from each patch. 

There is a subtlety here.
As discussed in \Sec{sec:signal1}, the resonance has a time duration of $L_{\rm coh}$ due to the finite frequency width of a resonance. Thus, not all resonance stimulates axion decays simultaneously. It is complicated to account for the fraction of GWs participating in the stimulation at each moment. But it is the original chirping GW which is largest and dominantly stimulating the axion decay; while at later time of the propagation, when the enhanced signal grows larger than the original chirping one, this issue becomes more relevant.

Rather than figuring out an accurate method, we estimate the range of the maximum and minimum possible summation. The enhancement in one patch is $1+\delta(f)$ from \Eq{eq:F} (and \Eq{eq:Fpatch}), where $\delta \ll 1$ is peaked at the central resonance frequency $f_0$. What is the enhancement after passing $N$ patches? The maximum summation assumes that all the axion signals from one patch contribute to the stimulation in the next patch, yielding the maximum total enhancement
\beq
(1+\delta(f))^N \,\approx\, e^{N\delta(f)} \qquad \textrm{(maximum sum)}.  \label{eq:maxsum}
\eeq
On the other hand, the minimum summation assumes no axion signals but only original GW stimulates in the next patch. This yields the minimum total enhancement
\beq
1+N\delta(f) \qquad \textrm{(minimum sum)}.  \label{eq:minsum}
\eeq
We use these two estimations to obtain an uncertainty band of our estimation (see \Fig{fig:bounds}, for example). A more realistic summation is likely to be between them.

\medskip

In both cases, the summation depends on the $N \delta(f)$. As $\delta$ depends linearly on $\rho$ ($\delta \propto \gamma^2 \propto \rho$), we can use the line-averaged density along the line of sight (LOS), for each coherent patch. For a LOS toward outside the galactic halo, the following line-averaged density is obtained 
\begin{equation}
\bar{\rho}_{\rm LOS} \= \frac{\int^\text{100 kpc}_\text{8 kpc} \rho(r) dr}{(\text{100 kpc}-\text{8 kpc})} \, \approx \, 0.04 \, \text{GeV/cm}^3  \label{eq:rhoave}
\end{equation}
for both NFW and Burkert profiles $\rho(r)$ of the MW dark matter halo, where 100 kpc is the assumed halo radius and 8 kpc is our distance from the MW center. We have taken best-fit parameters for both profiles from \cite{Nesti:2013uwa}. We have checked that $\int^\text{100 kpc}_\text{8 kpc} \rho(r) dr \simeq 0.98 \int^\infty_\text{8 kpc} \rho(r) dr$ for both profiles, confirming that the 100 kpc radius is sufficient. We use this average density value for $\rho_a$ in our numerical study. 

\medskip

Last, the factor  $N \delta$ makes the importance of finite coherence in yet another manifest way. From \Eq{eq:gamma} and \Eq{eq:Fpatch}, we have $\delta_{\rm patch}(f) \propto \rho \ell^4 m_a^2 / \Delta v^2 \times \text{sinc}^2(\epsilon/2 \Delta v)$. For the travel through the MW axion halo of size $R$, there are $N = R m_a \Delta v$ number of coherent patches. So the whole enhancement depends on the combination 
\beq
(N \delta(f))_{\rm halo} \, \propto \, \frac{R \rho_a \ell^4 m_a^3}{\Delta v} \times \text{sinc}^2\left(\frac{\epsilon}{2 \Delta v}\right).  \label{eq:ndelta}
\eeq
The linear dependence on the $R$ and $\rho$ is reasonable, and the overall dependence on $1/\Delta v$ implies that the enhancement is greater for the longer coherence from the smaller velocity dispersion. Thus, the effect of finite coherence indeed suppresses the size of the enhancement while broadening the frequency width.

\section{LIGO bounds and prospects} \label{sec:bound}

We use 11 GWs observed in LIGO O1 and O2 to obtain constraints on the axion coupling. As discussed in \Sec{sec:signal1}, every observed GWs will exhibit a common resonance peak due to the MW halo. In this section, assuming that the correlation of GWs can be made to find the common peak, we focus on individual GW properties in estimating the LIGO sensitivities.

\subsection{Detection criteria} \label{sec:criteria}

We measure the likelihood $L$ of the existence of a resonance peak using the peak strength as
\beq
-2 \ln L \= \chi^2 \,\equiv \, \sum_i (\Delta {\rm SNR}_{{\rm peak}, i})^2,
\eeq
where $i$ is summed over all observed GWs. The peak strength $\Delta {\rm SNR}_{\rm peak}$ is defined as the SNR in the resonance region ($-2 \pi \Delta v \leq \epsilon \leq 2 \pi \Delta v$ from \Eq{eq:Fpatch}) subtracted by the original SNR of the chirping GW; this roughly measures the significance of the deviation from smooth chirping. The $\Delta {\rm SNR}_{\rm peak}$ can be calculated from \Eq{eq:defF} or by multiplying \Eq{eq:maxsum} or \Eq{eq:minsum} to the original waveform.

We require the log-likelihood to be larger than 100: 
\beq
\chi \geq 100 \qquad \textrm{(detection criteria)}.
\eeq
This is the only requirement in our simplified analysis. Although this simple requirement can be mimicked by a strong peak in single GW, in real analysis the correlation of all the GWs (about the resonance shape in both frequency and time domain) must be made for further consistency. Assuming that such correlation can be made, we use the requirement $\chi \geq 100$ to estimate the LIGO bounds and prospects. 

The 100 is arbitrary but conservative requirement. The original SNR in the resonance region is $\sim {\cal O}(1)$ (for the 11 LIGO observations). Even if a somewhat larger frequency bin is used, $\Delta$SNR$_{\rm peak} \gtrsim 10$ might be good enough to be confidently detected; the fractional measurement uncertainty of the overall amplitude estimated by the Fisher information matrix is $\sim$1/SNR~\cite{Cutler:1994ys} so that our requirement is well above this sensitivity. We want to be conservative as real analysis including matched filtering and correlation may bring additional uncertainties. But the conservative estimation can be good enough because the signal strength $\Delta$SNR$_{\rm peak}$ depends on $\ell$ strongly ($N\delta \propto \gamma^2 \propto \ell^4$ from \Eq{eq:ndelta}), thus a mild improvement on the requirement does not bring large improvement on the $\ell$ bound. Therefore, while encouraging a more dedicated analysis, we are content with estimating conservative bounds and prospects based on our simplified analysis; see \Sec{sec:results} for the results and \Sec{sec:timedelay} for other realistic aspects.

\subsection{Results} \label{sec:results}

\begin{figure}[t] \centering
\includegraphics[width=0.48\textwidth]{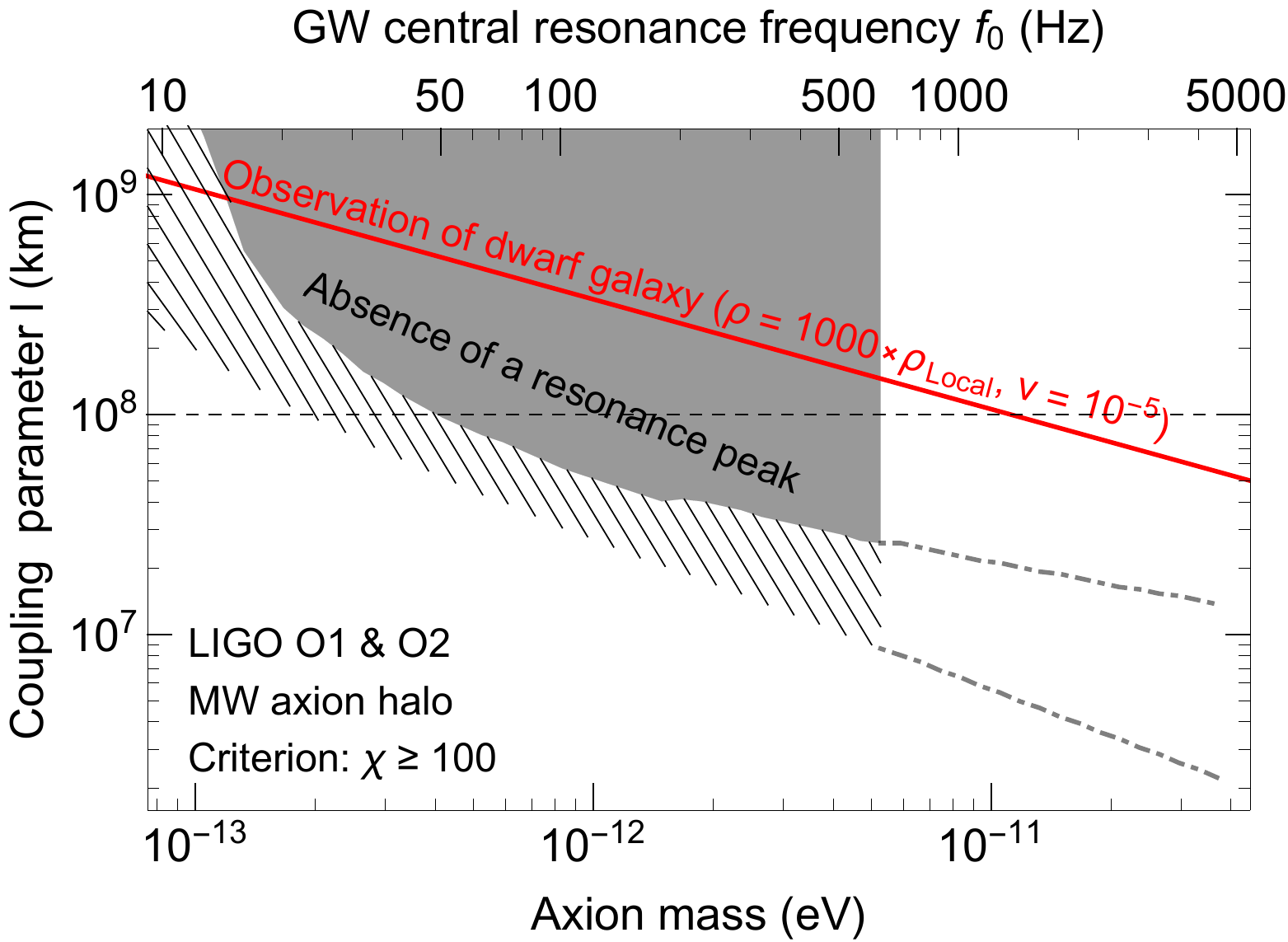}
\caption{
The upper limit on the axion Chern-Simons coupling $\ell$, assuming the absence of a resonance peak in the 11 GW observations at LIGO O1+O2. The gray shaded region is excluded. The hatched band indicates ambiguities in summing up effects from multiple coherent patches in a halo (\Sec{sec:addup}). The dot-dashed extensions are the bounds that can be achieved with one more NS-NS observation (to be correlated with the one existing observation).
The red solid is the bound from the existence of dwarf galaxies (\Sec{sec:signal2}). 
The horizontal dashed is the established bound from the Gravity Probe B~\cite{AliHaimoud:2011fw}.
} 
\label{fig:bounds}
\end{figure}

In \Fig{fig:bounds}, we show the LIGO bound on the axion Chern-Simons  coupling $\ell$ as a function of the axion mass $m_a$ (the corresponding peak GW frequency $f_0$ is shown on the upper horizontal axis). The gray shaded region is excluded, from the absence of a resonance peak in the 11 LIGO observations so far (Signal 1 in \Sec{sec:signal1}); each GW is considered up to its innermost stable circular orbit. This region is obtained by the most pessimistic summation of multi-patch effects as in \Eq{eq:minsum}. The hatched region indicates ambiguities in the summation method; this is the region that could be excluded if a somewhat more optimistic summation can be used. This region extends to the lower range of $\ell$ obtained by the most optimistic summation in \Eq{eq:maxsum}. A more realistic bound may lie somewhere in this band (\Sec{sec:addup}). The dot-dashed extension of the bounds are the expected bounds with one more NS-NS observation so that a correlation can be made with existing NS-NS data in the highest frequency range ${\cal O}(1000)$ Hz; heavier binaries merge at lower frequencies. The existing bound from Gravity Probe B satellite measurement of the frame dragging effect~\cite{AliHaimoud:2011fw} is shown as the horizontal dashed. The bound from the existence of dwarf galaxies (imposing that such systems are not exploded by resonant enhancement) is shown as the red solid (Signal 2 in \Sec{sec:signal2}). This is weaker than the previous two.

The LIGO bound from the absence of a peak is stronger than the existing established bound, at least for the axion mass range $5 \ \times 10^{-13} \ \text{eV}\lesssim m_a \lesssim 5\times 10^{-12} \ \text{eV}$. The bound can be stronger if a more aggressive summation can be used, and the heavier mass range up to $m_a \simeq 5 \times 10^{-11}$ eV can be constrained if more NS-NS mergers are observed, as discussed.

How will the bound improve with more data and smaller noise? For example, a 10 times smaller noise (achievable with, e.g., Einstein Telescope) will enhance SNR by 10 and observe $10^3$ times more GWs, yielding $\approx (10 \times \sqrt{1000})^{1/4} \simeq 4.2$ times stronger bound on $\ell$. Similarly, $n$ times smaller requirement on $\chi$ means $n^{1/4}$ times stronger bound on $\ell$. The measurement of lower frequency range from future GW detectors can also provide new constraints on the lower range of $m_a$. 

One can also note that the bound becomes stronger for the heavier axion. This is basically because $\gamma \propto m_a$ for a given axion energy density $\rho_a$ (see \Eq{eq:gamma0}), giving $(N \delta)_{\rm halo} \, \propto m_a^3$ as in \Eq{eq:ndelta}. This strong dependence on $m_a$ overcomes the frequency dependences of the noise curve and chirping GW spectrum; but slight mass dependence of the bound comes from these.

In all, LIGO is potentially able to improve the bound on the axion-gravity Chern-Simons coupling. We encourage a careful reanalysis of the currently available data.

\subsection{Time-delay of a resonance from dispersion} \label{sec:timedelay}
The resonant enhancement also modifies the group velocity of a resonance peak, delaying the arrival of  the peak relative to other frequency parts of chirping GW. As the original chirping GW has almost one-to-one correspondence between the frequency and arrival time, this time-delay produces an observable change of the time-domain waveform of the GW. 

The dispersion relation can be obtained from the correction term in \Eq{eq:psi} and ansatz (\ref{eq:ansatz}) for the case of small enhancement ($\mu t \ll 1$) in the vicinity of $\epsilon=0$ as 
\begin{equation}
\omega(k) \= k - \left(k-\frac{m_a}{2}\right) \left(\frac{m_a}{2}\gamma t\right)^2. \label{eq:dispersion}
\end{equation}
This means that $\omega = k$ at $t=0$ (not enhanced yet) but starts to deviate from $k$ as GW propagates through a coherent patch. Note that the dispersion is parity independent; see \Sec{sec:parityv} for usual parity-dependent dispersion. For $k>m_a/2$, $\omega$ decreases from $k$ toward $m_a/2$, and opposite for $k<m_a/2$. As $\mu t$ grows larger than 1, referring back to more general equation \Eq{eq:sol}, we find that the phase converges to some constant which implies (by ansatz \Eq{eq:ansatz}) $\omega = m_a / 2$ regardless of $k$. This behavior is approximately understood because the GW produced from axion decays has $\omega = m_a/2$ by the energy conservation, while its spatial mode is determined by initial chirping GW with the wavenumber $k \ne m_a/2$. As the enhancement grows, $\omega = m_a/2 =k$ dominates a whole GW.

Back to \Eq{eq:dispersion} with $\mu t \ll 1$, the group velocity of the axion signal $v_g = d\omega/dk$ at $\epsilon=0$ is 
\beq
v_g \= 1- \left(\frac{m_a}{2} \gamma t \right)^2.
\eeq
This again means that $\omega = k$ and $v_g=1$ at $t=0$ (not enhanced yet) starts to deviate with $t$. The group velocity is less than 1 so that the axion signal arrives later than the chirping GW. This could complicate the search because too large time delay will conceal the correlation between the appearance of axion signal and the arrival of chirping GW. Thus we estimate the time delay.

\begin{figure}[t] \centering
\includegraphics[width=0.48\textwidth]{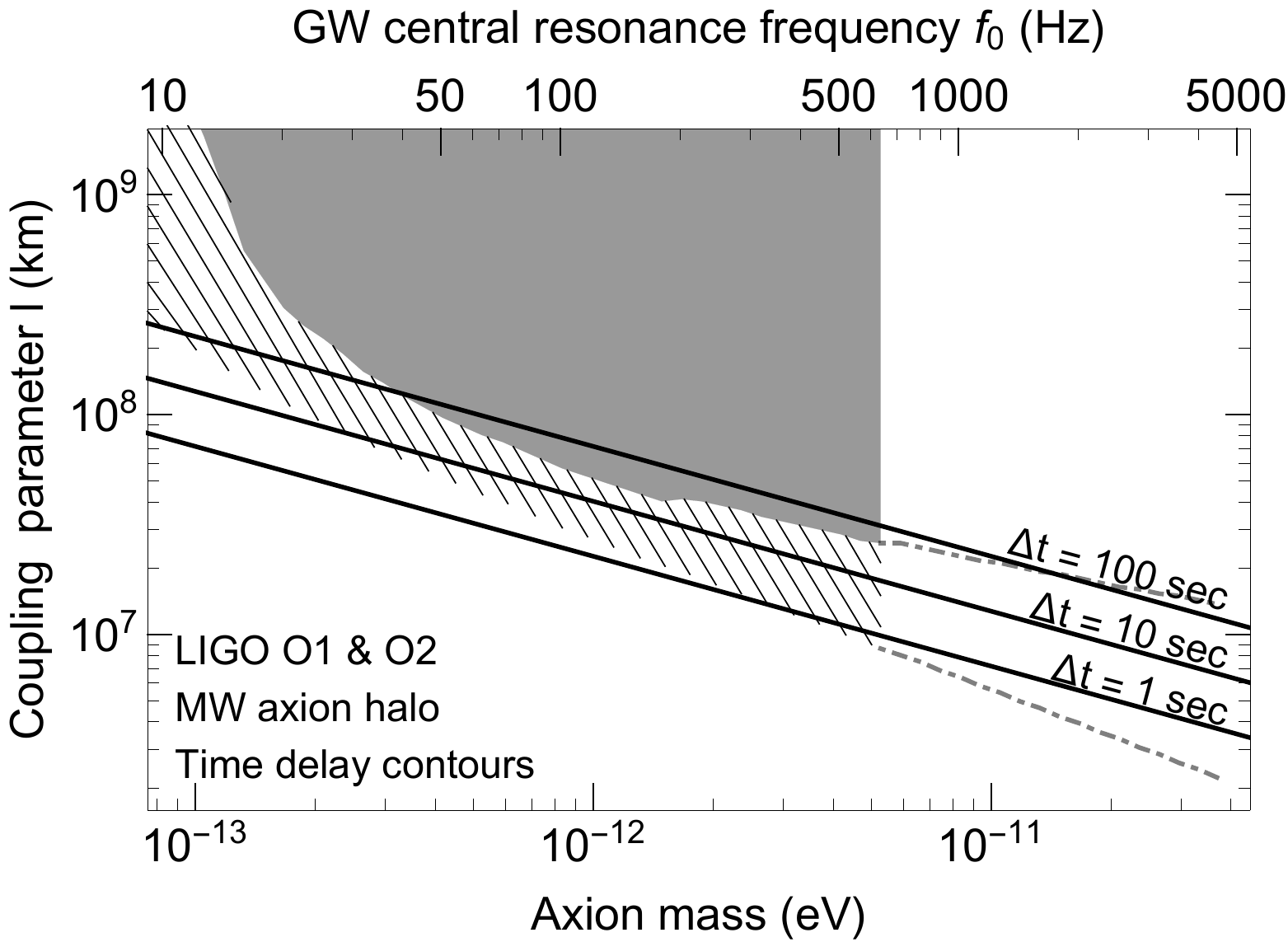}
\caption{
The contours (solid) of the arrival-time delay of a resonance peak with respect to the original chirping GW, induced from the propagation through a 100 kpc axion halo. The overlaid are the exclusion plots in \Fig{fig:bounds}.
} 
\label{fig:timedelay}
\end{figure}

The average group velocity during the propagation through one coherent patch (and this is the average group velocity in the galactic axion field) is given by 
\beq
\bar{v}_g \= 1-\frac{1}{3}\left(\frac{\gamma}{2\Delta v} \right)^2. \label{eq:vgavg}
\eeq
\Eq{eq:vgavg} gives the time delay of the resonance peak with respect to the chirping GW. In \Fig{fig:timedelay}, we plot this time-delay contours. In the parameter space that can be probed at LIGO, the time delay is 1 -- 100 seconds, which is also about the duration of a resonance $\sim 1/m_a \Delta v =$ 1 -- 100 seconds. This time scale is, however, longer than the typical duration (seconds or less) of chirping GWs in the LIGO band. Thus, we need to include longer time-series of data in order  to capture the peak which may not be much time-overlapped with the chirping GW. 

%

\subsection{Similar bounds on the axion-photon coupling} \label{sec:axion-photon}

We briefly comment on the axion-photon coupling. Since the solution of the coupled EM wave equations is in the same form as \Eq{eq:sol}~\cite{Arza:2018dcy, Wang:2020zur}, we can readily apply the same analysis done here to the photon case. The signal would be the extragalactic EM waves with a common peak.  
By simply requiring the maximum total enhancement of a single good EM signal to be greater than 10 (as light measurements are more precise),
we estimate the bound on the axion-photon coupling to be $g_{a\gamma \gamma} \lesssim 10^{-8} \text{--} 10^{-2} \ \text{GeV}^{-1}$ for the axion mass range $10^{-11} \text{--} 1 \ \text{eV}$. This is similar or slightly weaker than the laboratory bounds, while much weaker than the Helioscope bound by about 2 -- 8 orders of magnitudes~\cite{Irastorza:2018dyq}. We defer more detailed analysis and comparison to a future project.

\section{Discussions}  \label{sec:resonance}

\subsection{Energy conservation and axion backreaction} \label{sec:energyconservation}

The energy conservation implies that the amplitude of the axion field should decrease as the GW amplitude is enhanced. However, dark matter energy density is much greater than any reasonable GW energy density. We can estimate the chirping GW energy density as the following. GW150914 emitted $\sim 3 \Msun$ of energy at 400 Mpc, and by assuming all the energy was released in the last 0.1 second of chirping, we have $\rho_{\rm GW} = 60 \ \text{eV/cm}^3$. This is incomparably smaller than the dark matter energy density $\bar{\rho}_{\rm DM} = 0.04 \ \text{GeV/cm}^3$ in \Eq{eq:rhoave}. Thus, we can assume that axion fields do not decrease in our work.  

But if somehow energies of both waves become similar, the coupled wave equations describe the energy transfer between them through the time-evolution of both amplitudes.
For example for the EM wave case, Eq.(8) and Fig.2 of \cite{Arza:2018dcy} show such time-variation. Back to a general point of view, the absence of explicit time dependence of the Lagrangian guarantees the energy conservation for a dynamically-evolving axion field. After all, the backreaction of axion fields will stop the exponential growth of GWs (explosion) at some point.

\subsection{Stimulated axion decay rate} \label{sec:decayrate}

We obtain another insight on the stimulated decay by calculating the axion decay rate from the energy gain of the GW, which equals to the energy loss of the axion. Since the spatially averaged energy density of GW is given by $\langle \rho_{\rm GW} \rangle = \omega_{\rm GW}^2 (|\tilde{h}^{(R)}|^2+|\tilde{h}^{(L)}|^2) / 64\pi G$, the energy density gain of the forward and backward waves are 
\begin{eqnarray}
\Delta \langle \rho_{\rm GW} \rangle_F &=&
\langle \rho_{\rm GW} \rangle_0 \left(F^2(t)-1\right) \nonumber \\
&\approx&\langle \rho_{\rm GW} \rangle_0 \times \gamma^2 \left(\frac{m_a}{2}\right)^2 \frac{\sin^2\left[\left(\omega_{\rm GW} - \frac{m_a}{2}\right) t\right]}{\left(\omega_{\rm GW} - \frac{m_a}{2}\right)^2} \nonumber \\
&=& \Delta \langle \rho_{\rm GW} \rangle_B,
\end{eqnarray}
where the last equality is due to the momentum conservation (this can also be explicitly derived from \Eq{eq:solback}). 

From the energy density loss of the axion $2\times \Delta \langle \rho_{\rm GW} \rangle_F$, we obtain the decay rate of the axion as (for $\mu t \ll 1$; otherwise, the rate increases exponentially)
\begin{eqnarray}
P_{\rm decay}(t) &=& \frac{-\Delta \rho_a}{\rho_a}  \\
&=&\langle \rho_{\rm GW} \rangle_0 \times \frac{1}{2}\pi G l^4 m_a^4  \frac{\sin^2\left[\left(\omega_{\rm GW} - \frac{m_a}{2}\right) t\right]}{\left(\omega_{\rm GW} - \frac{m_a}{2}\right)^2}. \nonumber
\end{eqnarray}
As it should be, this is proportional to the energy density of the GW and independent on the axion energy density. Remarkably, the form of $\sin^2((\omega-\omega_0)t)/(\omega-\omega_0)^2$ is nearly identical to the probability of stimulated emission in quantum mechanics \cite{griffiths2010introduction}. This consideration supports the physical picture of the resonant enhancement as the stimulated decay.

\subsection{Effective `graviton' mass}

Even without the axion-gravity Chern-Simons  coupling, the GW experiences a dispersion due to the intervening mass density, similarly to the photon's plasma mass in the electron medium. Following the EM wave case in \cite{Hertzberg:2018zte}, we check that such effect is negligible for the GW.

The GW refractive index is given by $n = 1+2\pi G \rho / \omega^2$~\cite{Peters:1974gj}. This gives the dispersion relation $\omega^2 = k^2 - 4\pi G \rho$. The ratio of the matter-induced dispersion to the effect of gravitational Chern-Simons coupling in \Eq{eq:Mathieu} is
\beq
\frac{4 \pi G \rho}{m^2 \gamma} \= 6.3\times 10^{-50} \times \sqrt{\frac{\rho}{\text{GeV/cm}^3}} \times \left(\frac{m}{\text{eV}}\right)^{-3} \times \left(\frac{\ell}{\text{km}}\right)^{-2}
\eeq
at the resonance. This is an incredibly small number; for instance, $\rho = 1 \ \text{GeV/cm}^3$, $m = 10^{-12} \ \text{eV}$, and $\ell = 10^8 \ \text{km}$ give $6.3\times 10^{-30}$. Thus, the dispersion of GWs due to intergalactic matter can be ignored.

\subsection{Axions in the source galaxy and intergalactic region} \label{sec:IGM}

The sharp peak was unambiguously associated with the axion signal because every GWs will exhibit a common peak from the propagation through the Milky-Way axion halo. What about axions in other galaxies (in particular, the one that hosts the source of the GW) and in intergalactic region?

First, the cosmological redshift of a source galaxy varies among different sources. So does the observed peak frequency. Such peaks may not be confidently identified.

The effect from intergalactic dark matter may not be strong enough due to low density and redshift. While a half of total dark matter resides in the intergalactic region, the density there which can be estimated as $\Omega_{\rm DM} \rho_{c} \sim 10^{-6}$ GeV/cm$^3$ is $10^{-4}$ times smaller than  $\bar{\rho}_a$ in the MW halo (\Eq{eq:rhoave}). Thus, the enhancement factor $N \delta \propto R \rho_a / v$ (\Eq{eq:ndelta}) either needs more than a few Gpc propagation ($10^4$ times longer than the MW halo size) or much smaller velocity dispersion, to produce similar size of total enhancement. The small dispersion is unlikely; for example, the Local Group velocity with respect to the CMB and the escape velocity of the galaxy are all ${\cal O}(100)$ km/s $\gtrsim 10^{-3}$. In addition, continuously varying redshift through the intergalactic region will further hinder the generation of a sharp and large axion signal peak. Thus, we ignore the intergalactic contributions.

\section{Corollary: Absence of parity-violation observables on the chirping GW} \label{sec:parityv}

Many previous works have studied the parity-violation signals in photons due to the axion Chern-Simons  coupling.
However, our solution shows that the parity-violation is not observable in the resonance regime of the chirping GW. They are not in contradiction as we explain in this section.

The parity violation in previous works are manifest in two ways: one is through the dispersion relation~\cite{Yoshida:2017ehj, Ivanov:2018byi, Harari:1992ea, Fedderke:2019ajk, Lue:1998mq, Chu:2020iil, Sigl:2018fba} and the other through the enhancement~\cite{Yoshida:2017ehj, Yoshida:2017cjl}. For our case, the parity-dependent dispersion is absent because we consider the resonance regime ($k \approx m_a/2$), and the parity-dependent enhancement is absent because we consider the forward propagation of waves (not stochastic waves).

First, the parity-dependent dispersion is obtained from the wave equation in the form of \Eq{eq:Mathieu} as
\beq
\omega(k)^2 \= k^2 + 2 \lambda^{(s)} m_a k \gamma \sin(m_a t).
\eeq
This shows the usual parity-dependent ($\lambda^{(s)}$-dependent) dispersion relation, making the phase velocity deviate from $c$ and oscillate oppositely for opposite polarizations. But the deviation (the second term) oscillates in time with the frequency $1/m_a$. In previous works with $k \gg m_a/2$, this oscillation was much slower than the high frequency of photons. But in our case in the resonance regime, they are comparable ($k \approx m_a/2$) so that the deviation almost averages out in one period of the GW/photon. Instead, near the resonance, there arises the dispersion relation which is parity independent, as discussed in \Sec{sec:timedelay}.

Second, the usual parity-dependent enhancement arises from the time evolution of spatial Fourier modes, $\tilde{h}(k, t) = \int h(x, t) e^{- i k x}$ ~\cite{Yoshida:2017ehj, Yoshida:2017cjl}. But these spatial modes are the sum of forward and backward propagating waves. By separating the propagation direction, as in our solution \Eq{eq:sol} and \ref{eq:solback}, we find that the parity-violation exists only in backward waves due to the initial condition $h_B=0$. This effect is not observable in our case because what we observe is only $h_F$. The previous solutions are suitable for stochastic backgrounds, like CMB or stochastic GWs, where waves with all directions are mixed up. Such waves can exhibit the parity violation as polarization-dependent enhancements, as studied in previous works.

\section{Conclusions}  \label{sec:conclusion}

We have shown that the LIGO observation of chirping GWs can constrain the axion-gravity Chern-Simons coupling, through a resonance peak of the GW induced by the coherently oscillating axion dark matter field. As all the observed GWs will have a peak with common properties (frequency, duration, and height), the correlation among them can confidently detect or reject the peak. We have found that 11 GW observations at LIGO O1+O2 can already provide the strongest bound on the coupling, at least for $m_a = 5 \times 10^{-13} \sim 5 \times 10^{-12}$ eV (see \Fig{fig:bounds}). With more LIGO observations, the range can be extended and the bound can be stronger. A careful reanalysis of existing data is encouraged.

The resonance phenomenon is essentially the stimulated decay of the axion. Not only does the resonance condition $f_0 \simeq m_a/2$ support this particle-like interpretation, but also the decay probability estimated from the energy gain and loss of the fields agrees with the quantum mechanical description of stimulated emissions and absorptions. This is remarkable as we have never quantized these waves. 

The finite coherence of the axion field determines the resonance (axion signal) shape in large part. First, it suppresses the height of the resonance peak in the chirping GW spectrum, while broadening the peak width. The  broadening in the frequency domain also makes the signal persist as long as the size of a coherent patch. The resonance-produced axion signal is also time-delayed compared to the original chirping GW, and this time-delay is also affected by the finite coherence. Resonance searches must account for these effects.

A proper ansatz treating forward and backward-going waves separately is crucial for our work. It is because only the forward-going chirping GW can be observed. This is different from the stochastic background of GWs and CMB, where waves with all directions are mixed up. As a consequence, our solution does not exhibit the parity violation from the axion Chern-Simons coupling in the forward wave at the linear order, but this is not in contradiction with previous studies.

Last but not the least, the resonant effect can sometimes become so efficient that an axion substructure may not exist today. This happens when the axion structure has small velocity dispersion (hence, long coherence) and high density. The axion minicluster is one example that might have exploded by today, but the coherent axion field virialized with a whole galaxy does not explode given the current bound on the coupling. Although we assumed that the signal of explosion had diffused away, it would be interesting to study if any observable signals remain.

In all, we have studied one way to probe and constrain the axion-gravity Chern-Simons coupling, which is generic and well motivated. A careful reanalysis of LIGO data may provide one of the strongest constraints on this coupling. Various other types of axion-gravity couplings may also be probed in a similar way.

\begin{acknowledgments}
Authors would like to thank Han Gil Choi, Kwang Sik Jeong, Hyung Do Kim, Hyungjin Kim, Ji-hoon Kim, Chang Sub Shin, Takahiro Tanaka for useful discussions.
SJ and THK are supported by the NRF of Korea under grants NRF-2019R1C1C1010050, 2015R1A4A1042542 and SJ also by POSCO Science Fellowship. 
JS is supported in part by JSPS KAKENHI Grant Numbers JP17H02894, JP17K18778, JP15H05895, JP17H06359, JP18H04589.
YU is supported by Grant-in-Aid for Scientific Research on Innovative Areas under Contract No.~18H04349, Grant-in-Aid for Scientific Research (B) under Contract No. 19H01894, and the Deutsche Forschungsgemeinschaft (DFG, German Research Foundation) - Project number 315477589 - TRR 211.
JS and YU were also supported by JSPS Bilateral Joint Research Projects (JSPS-NRF collaboration) ``String Axion Cosmology'' and benefitted from discussions during the YITP workshop YITP-T-19-02 on ``Resonant instabilities in cosmology''.
The research of SJ and YU was supported in part by the National Science Foundation under Grant No. NSF PHY-1748958.
\end{acknowledgments}

\appendix

\section{Solving wave equation through the Mathieu equation} \label{app:Mathieu}

In this appendix, we solve \Eq{eq:waveeq} by another method. By expressing $h_{ij}$ in \Eq{eq:waveeq} as a Fourier transform, the wave equation for each Fourier amplitude becomes \cite{Yoshida:2017cjl}
\begin{equation}
\ddot{\tilde{h}}^{(s)} + \frac{4 \lambda^{(s)} \gamma \cos(m_a t)}{1+ 4 \lambda^{(s)} \frac{k}{m_a} \gamma \sin(m_at)} k \dot{\tilde{h}}^{(s)} + k^2 \tilde{h}^{(s)} = 0, \label{eq:waveeqoriginal}
\end{equation}
where $\tilde{h}^{(s)}$ represents the spatial Fourier mode with wave number $k$ (so different from the amplitude $h_{F/B}^{(s)}$ appearing in \Eq{eq:ansatz}; $\tilde{h}^{(s)}$ contains the full oscillation part $e^{-i \omega t}$).
Since we only want to see the leading order, we rewrite the equation as
\begin{equation}
\ddot{\tilde{h}}^{(s)} + 4 \lambda^{(s)} \gamma \cos(m_a t) k \dot{\tilde{h}}^{(s)} + k^2 \tilde{h}^{(s)} = 0. 
\end{equation}
By following the transformation in \cite{Chu:2020iil} with cosmic expansion neglected, we define $\psi$ as
\begin{equation}
\tilde{h}^{(s)} = e^{-2 \int \lambda^{(s)}\gamma k \cos(m_a t) dt} \Psi, \label{eq:psidef}
\end{equation}
and we have
\begin{equation}
\ddot{\Psi} + \left(k^2 + 2 \lambda^{(s)} m_a k \gamma \sin(m_a t)\right)\Psi = 0 \label{eq:Mathieu}
\end{equation}
which is the ordinary Mathieu equation. The exponential factor in \Eq{eq:psidef} is a non-resonant term since the exponent oscillate with small amplitude. Any analytic method solving the Mathieu equation tracks only the resonant term, so this factor does not affect the result. Also physically, this factor will be canceled out in average, due to its dependence on the relative phase between the axion field and the gravitational wave (for such cases, the argument of cos's and sin's should have a constant phase term, like $mt+\psi_0$).

To solve \Eq{eq:Mathieu} by the two variable expansion method \cite{kovacic2018mathieu, rand2012perturbation} we look at the solution behavior near the resonance at $k=m_a/2$. To do this, we use $(4\gamma)$ for an expansion parameter, and the expansion will done up to the first order. The two variables in the expansion are the ordinary time $\xi = t$ and the slow time $\eta = 4 \gamma t$. $\xi$ is the time scale of wave oscillation, while $\eta$ is the time scale of amplitude change. We regard $\Psi$ as a function of the two \textit{independent} variables, as $\Psi = \Psi(\xi, \eta)$. And the time derivative operator becomes
\begin{equation}
\frac{d}{dt} = \frac{\partial}{\partial \xi} + 4\gamma \frac{\partial}{\partial \eta}. \label{eq:ddt}
\end{equation}
Similarily, $k$ and $\Psi$ are expanded as 
\begin{equation}
k = \frac{m}{2} \left[1+\epsilon_1 (4\gamma) + \epsilon_2 (4\gamma)^2 + \cdots \right] \label{eq:kexpand}
\end{equation}
and
\begin{equation}
\Psi(\xi, \eta) = \Psi_0(\xi, \eta) + (4\gamma) \Psi_1(\xi, \eta) + (4\gamma)^2 \Psi_2(\xi, \eta) + \cdots \ .\label{eq:psiexpand}
\end{equation}

By putting Eqs. (\ref{eq:ddt})--(\ref{eq:psiexpand}) into \Eq{eq:Mathieu}, we can obtain series of equations assorted by the order in $(4\gamma)$. The 0th order equation is
\begin{equation}
\frac{\partial^2 \Psi_0}{\partial \xi^2} + \left(\frac{m_a}{2}\right)^2 \Psi_0 = 0, \label{eq:0th}
\end{equation}
and the 1st order equation is (note that $\xi = t$)
\begin{eqnarray}
&&\frac{\partial^2 \Psi_1}{\partial \xi^2} + \left(\frac{m_a}{2}\right)^2 \Psi_1 \nonumber \\
&&= -2 \frac{\partial^2 \Psi_0}{\partial \xi \partial \eta} - \lambda^{(s)} \frac{m_a^2}{4} \sin(m_a\xi) \Psi_0 - \frac{m_a^2}{2}\epsilon_1 \Psi_0. \qquad \label{eq:1storiginal}
\end{eqnarray}
\Eq{eq:0th} gives $\Psi_0$ in the form of
\begin{equation}
\Psi_0 = A(\eta) \cos(\frac{m_a}{2}\xi) + B(\eta) \sin(\frac{m_a}{2}\xi), \label{eq:psi0}
\end{equation}
and putting this into \Eq{eq:1storiginal} with using trigonometric identities gives
\begin{eqnarray}
&&\frac{\partial^2 \Psi_1}{\partial \xi^2} + \left(\frac{m_a}{2}\right)^2 \Psi_1 \nonumber \\
&&= \left(m_a A'(\eta) - \lambda^{(s)} \frac{m_a^2}{8} A(\eta) - \frac{m_a^2}{2}\epsilon_1 B(\eta) \right)\sin(\frac{m_a}{2}\xi) \nonumber \\
&& \quad + \left(-m_a B'(\eta) - \lambda^{(s)} \frac{m_a^2}{8} B(\eta) - \frac{m_a^2}{2}\epsilon_1 A(\eta) \right)\cos(\frac{m_a}{2}\xi) \nonumber \\
&& \quad + \cdots \label{eq:1stresonance}
\end{eqnarray}
where the higher frequency terms in RHS whose resonance appear only in higher orders are not shown. This gives the slow flow equations for $A$ and $B$ as
\begin{equation}
\frac{d}{d\eta}
\begin{pmatrix}
A(\eta) \\
B(\eta)
\end{pmatrix}
=
\begin{pmatrix}
\lambda^{(s)} \frac{m_a}{8} & \frac{m_a}{2} \epsilon_1 \\
-\frac{m_a}{2} \epsilon_1 & -\lambda^{(s)} \frac{m_a}{8}
\end{pmatrix}
\begin{pmatrix}
A(\eta) \\
B(\eta)
\end{pmatrix}. \label{eq:matrixform}
\end{equation}
We find a solution in the form of
\begin{equation}
\begin{pmatrix}
A(\eta) \\
B(\eta)
\end{pmatrix}
=
\begin{pmatrix}
a \\
b
\end{pmatrix}
e^{\nu \eta}. \label{eq:matrixansatz}
\end{equation}
By putting \Eq{eq:matrixansatz} into \Eq{eq:matrixform}, we get 
\begin{equation}
\nu = \pm \frac{m_a}{8} \sqrt{1-16\epsilon_1^2}, \label{eq:nu}
\end{equation}
and
\begin{subequations}
\begin{equation}
\begin{pmatrix}
a \\
b
\end{pmatrix}_{+}
=
\begin{pmatrix}
1 \\
-\frac{\lambda^{(s)} - \sqrt{1-16 \epsilon_1^2}}{4\epsilon_1}
\end{pmatrix},
\end{equation}
\begin{equation}
\begin{pmatrix}
a \\
b
\end{pmatrix}_{-}
=
\begin{pmatrix}
-\frac{\lambda^{(s)} - \sqrt{1-16 \epsilon_1^2}}{4\epsilon_1} \\
1
\end{pmatrix}.
\end{equation} \label{eq:eigenvector}
\end{subequations} 

We put these to \Eq{eq:psi0} and recover the notations we used in the main paper. Since $\epsilon$ was used to denote the \textit{fractional} difference between $k$ and $m_a/2$, we have $\epsilon_1 = \epsilon / (4\gamma)$ (see \Eq{eq:kexpand}). Then, we have $\nu \eta = \pm \mu t$ and the solution we obtained is (recall $\xi = t$)
\begin{eqnarray}
\Psi(t) &=& C_1 \left(\cos(\frac{m_a}{2}t) -\beta^{(s)} \sin(\frac{m_a}{2}t)\right)  e^{\mu t} \nonumber \\
&& + C_2 \left(-\beta^{(s)} \cos(\frac{m_a}{2}t) + \sin(\frac{m_a}{2}t)\right)  e^{-\mu t}. \qquad \label{eq:sol1}
\end{eqnarray}
where
\begin{equation}
\beta^{(s)} \equiv \frac{\lambda^{(s)} - \sqrt{1-16 \epsilon_1^2}}{4\epsilon_1} \label{eq:betas}
\end{equation}
and $C_1$ and $C_2$ are arbitrary constants.

Then we extract the forward and backward waves from \Eq{eq:sol1} by expressing cosine and sine functions in terms of exponentials, as
\begin{eqnarray}
&&\Psi(t) \nonumber \\
&& = \left[D_1 (1+ i \beta^{(s)}) e^{\mu t} + D_2 (1- i \beta^{(s)}) e^{-\mu t} \right] e^{i \frac{m_a}{2} t} \nonumber \\
&& \quad + \left[D_1 (1- i \beta^{(s)}) e^{\mu t} - D_2 (1+ i \beta^{(s)}) e^{-\mu t}\right] e^{-i \frac{m_a}{2} t} \qquad \quad
\end{eqnarray}
where the arbitrary constants are redefined. Recalling that we started from \Eq{eq:waveeqoriginal} about the spatial Fourier mode $\sim e^{i k x}$, the $e^{i \frac{m_a}{2} t}$ part denotes the backward wave and the $e^{- i \frac{m_a}{2} t}$ part is for the forward wave. We now put the initial condition of vanishing backward wave at $t=0$. We first write the $D$ coefficients as $D_1 = D (1-i \beta^{(s)})$ and $D_2 = -D (1+i\beta^{(s)})$, and normalize by the initial amplitude of the forward wave $h_0 = 2D(1-(\beta^{(s)})^2)$. This gives
\begin{eqnarray}
\Psi(t) &=& h_0 \frac{1+ (\beta^{(s)})^2}{1-(\beta^{(s)})^2} \sinh(\mu t) e^{i \frac{m_a}{2} t} \nonumber \\
&& + h_0 \left[\cosh(\mu t) -i \frac{2 \beta^{(s)}}{1-(\beta^{(s)})^2} \sinh(\mu t) \right] e^{-i \frac{m_a}{2} t}. \nonumber \\
\end{eqnarray}
To further simplify the coefficients, we recall \Eq{eq:betas}. Since $\lambda^{(s)} = \pm 1$, we have
\begin{eqnarray}
\Psi(t) &=& h_0 \lambda^{(s)} \frac{\gamma}{\sqrt{\gamma^2 - \epsilon^2}} \sinh(\mu t) e^{i \frac{m_a}{2} t} \nonumber \\
&& + h_0 \left[\cosh(\mu t) -i  \frac{\epsilon}{\sqrt{\gamma^2-\epsilon^2}} \sinh(\mu t) \right] e^{-i \frac{m_a}{2} t}. \nonumber \\
\end{eqnarray}

This is the same solution in Eqs. (\ref{eq:sol}) and (\ref{eq:solback}), considering the phase factors in the ansatz, \Eq{eq:ansatz}. Since we assumed initial phase of the axion field to be zero, it does not appear here. Also, here the initial amplitude of the forward wave $h_0$ is assumed to be real. Thus, we have obtained identical solution via solving the Mathieu equation.


\end{document}